\documentclass[12pt]{iopart}
\usepackage{iopams,graphicx,epstopdf,cite,amssymb}
\usepackage{appendix,hyperref,xcolor}

\expandafter\let\csname equation*\endcsname\relax
\expandafter\let\csname endequation*\endcsname\relax
\usepackage{amsmath}

\let\origleft\left
\let\origright\right

\renewcommand{\left}{\mathopen{}\mathclose\bgroup\origleft}
\renewcommand{\right}{\aftergroup\egroup\origright}

\newcommand{\be}{\begin{equation}}
\newcommand{\ee}{\end{equation}}

\newcommand{\bef}{\begin{figure}}
\newcommand{\enf}{\end{figure}}
\newcommand{\p}{\mathbf{p}}
\renewcommand{\r}{\mathbf{r}}

\newcommand{\F}{\mathbf{F}}

\newcommand{\A}{\mathbf{A}}

\newcommand{\w}{\omega}

\begin{document}

\title{Spin-Orbit Larmor Clock for Ionisation Times in One-Photon and Strong-Field Regimes}
\author{Jivesh Kaushal$^{1}$, Felipe Morales$^{1}$, Lisa Torlina$^{1}$, Misha Ivanov$^{1,2,3}$ and Olga Smirnova$^{1}$}
\address{$^{1}$Max-Born Institute for Nonlinear Optics and Short Pulse Spectroscopy, Max-Born-Strasse 2A, D-12489 Berlin, Germany\\
$^{2}$Department of Physics, Imperial College London, South Kensington Campus, SW7 2AZ London, United Kingdom\\
$^{3}$Institute f{\"u}r Physik, Humboldt-Universit{\"a}t zu Berlin, Newtonstrasse 15, 12489 Berlin, Germany}

\begin{abstract}
Photoionisation is a process where absorption of one or several photons liberates an electron and creates a hole in a quantum system, such as an atom or a molecule. Is it faster to remove an electron using one or many photons, and how to define this time? Here we introduce a clock that allows us to define ionisation time for both one-photon and many-photon ionisation regimes. The clock uses the interaction of the electron or hole
spin with the magnetic field created by their orbital motion, known as the spin-orbit interaction. The angle of spin precession in the magnetic field records time. We use the combination of analytical theory and ab-initio calculations to show how ionisation delay depends on the number of absorbed photons, how it appears in the experiment and what electron dynamics it signifies. In particular, we apply our method to calculate the derived time delays in tunnelling regime of strong-field ionisation.
\end{abstract}
\submitto{\jpb}
\maketitle

\section{Introduction}

Quantum mechanics uses operators to predict an outcome of a measurement. 
However, some figures of merit,
e.g. the phase of quantum electromagnetic field or the times of quantum transitions are not associated with operators. Their measurement has to rely on the operational approach, i.e. a measurement protocol yielding a particular observable at the detector. Measuring ionisation times with attosecond accuracy has been the focus of several recent experiments \cite{Uiberacker2007,Cavalieri2007,Eckle2008,Schultze2010,Klunder2011,
Shafir2012}, using different measurement protocols in different ionisation regimes.

For one-photon ionisation the definition of ionisation time $\tau_{WS}$, referred to as Eisenbud-Wigner-Smith time, is established \cite{Wigner1955,Smith1960} and verified in the analysis \cite{Klunder2011,mivanov,joachim,Maquet,Maquet2} of recent experiments\cite{Klunder2011,Schultze2010}. This definition links $\tau_{WS}$ to the phase of the photoelectron wave-function $\phi$ through its derivative with respect to electron energy: $\tau_{WS}=-d\phi/dE$. \textsf{Establishing such a link between the classical concept of time and the parameters of a quantum wave-function in the regime of strong field ionisation is the goal of this paper.}

To define the ionisation time for one-photon and multiphoton 
ionisation regimes within the same protocol, we extend the idea of the Larmor clock, originally introduced in \cite{Baz1966} to define the time it takes an electron to tunnel through a barrier (see e.g. \cite{Landauer1994,Hauge,Steinberg}). The Larmor clock measures rotation of the electron spin in an external homogeneous magnetic field acting exclusively inside the barrier. The angle of rotation is the hand of the clock.
Here, we introduce the analogue of this clock, which is based on the spin-orbit interaction and is naturally built into many atoms. Physically, the spin-orbit interaction can be understood by considering an electron with angular momentum $l$ orbiting around the nucleus. In the reference frame associated with the electron, the nucleus rotates around it and creates
a current. The current creates the magnetic field. Precession of the 
electron spin in this field records time.

\section{Spin-orbit Larmor clock for one-photon ionisation: calibration of the clock} \label{sec:ii}

\begin{figure}
\begin{center}
\includegraphics[width=14cm]{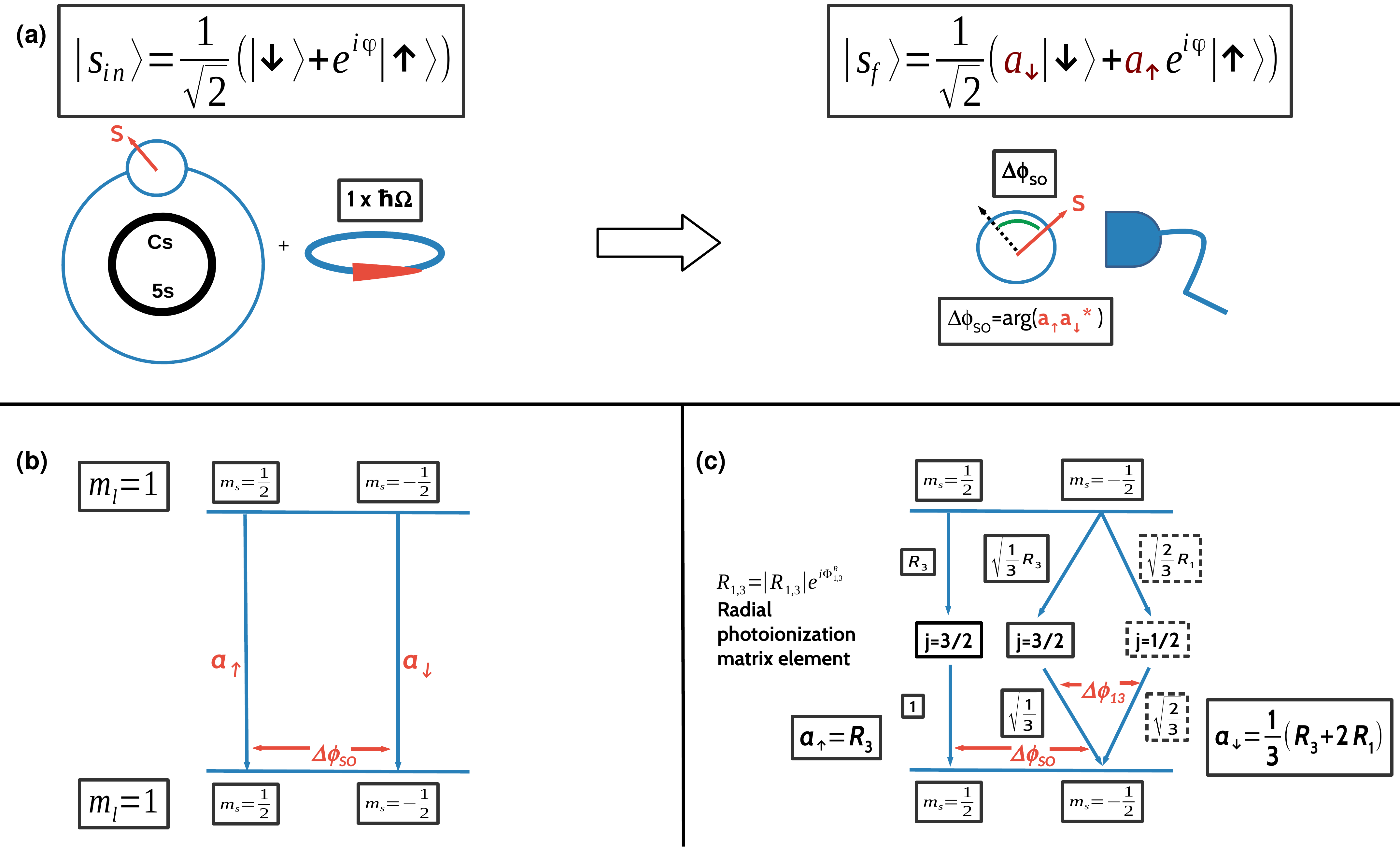}
\caption{Gedanken experiment for calibrating the spin-orbit Larmor clock in one-photon ionisation. (a) Cartoon of the experiment, for a Cs atom: a circularly polarised field removes an electron from the valence 5s shell, prepared in a coherent superposition of the spin-up and spin-down states. In the final state, the electron spin has rotated by angle $\Delta\phi_{SO}$. (b) The spin-orbit clock operating on the electron viewed as an interferometer, a simplified view. The two arms correspond to the spin-up and spin-down pathways, with phase difference $\Delta\phi_{SO}$; Formally, the interferometer describes the following interfering pathways, where $m_l'=1$, $m_s=m_s'=\pm 1/2$  correspond to two different arms of the interferometer: $\left\langle m_{l}',m_{s}',f \middle\vert \hat{R}\,\hat{\Xi} \middle\vert m_l=0,m_s,g \right\rangle\propto \left\langle m_{l}',m_{s}',f \middle\vert \hat{R} \middle\vert m_{l}=1,m_s,g \right\rangle$. Here $\hat{R}$ is the radial part of the dipole operator, $\hat{\Xi}$ is the angular part of the dipole operator, $g,f$ are the radial parts of the initial and final state
wavefunction. We have used that $\hat{\Xi}|m_l=0,m_s\rangle\propto  \vert m_{l} = 1,m_{s}\rangle $. (c) Detailed view of the spin-orbit interferometer. The spin-down path is itself a double arm, since the spin-down electron ($m_{s} = -1/2$) and the final orbital momentum $m_{l}=1$ can proceed via both $j=1/2$ and $j=3/2$ continua. The single (spin-up) arm and the double (spin-down) arm interfere in the final continuum state with $m_{l}=1$. Formally, it corresponds to the following interfering pathways: $\sum_{j,m_j} \left\langle m_l^{'},m_s^{'},f \middle\vert \hat{R} \middle\vert j,m_j\right\rangle \left\langle  j,m_j \middle\vert m_{l} = 1, m_{s}, g \right\rangle$, where $m_l'=1$, $m_s=m_s'=\pm 1/2$, correspond to two different arms of the interferometer.} \label{figure1}
\end{center}
\end{figure}

To illustrate how the clock works, consider a Gedanken one-photon 
ionisation experiment, where an $s$-electron is removed from an atom (e.g. Cs) by a right circularly polarised light field, see Fig.~\ref{figure1}(a). There is no spin-orbit interaction in the initial state--the ground state of a Cs atom. 
Spin-orbit interaction turns on upon photon absorption, since the electron angular momentum changes from $l=0$ ($s$-state) to $l=1$ with its projection on the laser propagation direction $m_l=1$. Thus, photon absorption turns on the Larmor clock: the electron spin starts to precess. Spin-orbit interaction is short-range, localised near the core. Once the freed electron leaves this area, spin precession stops and the clock turns off.

However, the original Larmor clock \cite{Baz1966} used 
{\it homogeneous} magnetic field and hence the clock hand rotated with constant speed. In our case, the spin-orbit interaction is inhomogeneous, requiring calibration of the clock: the mapping between the angle of rotation and the ionisation time. To calibrate the clock we consider rotation of electron spin during one-photon ionisation, where the ionisation time $\tau_{WS}$ is well established.

Let us prepare the electron in the initial spin-polarised state, $\vert s_{\rm in} \rangle = (1/\sqrt{2})[\vert -1/2 \rangle + e^{i\varphi}\vert 1/2 \rangle]$, with the phase $\varphi$ characterising the initial orientation of the spin in the polarisation plane, and calculate the angle of rotation of the electron spin in this plane during ionisation. The final state, for the final orbital momentum $m_l=1$, is $\vert s_{\text{fin}} \rangle=(1/\sqrt{2})[a_{\downarrow}\vert -1/2 \rangle + a_{\uparrow} e^{i\varphi} \vert 1/2 \rangle]$, where the $a_{\uparrow}$ and $a_{\downarrow}$ are the ionisation amplitudes for the spin-up and spin-down ionisation pathways.
The spin has rotated by the angle $\Delta\phi_{SO}=\arg[a_{\uparrow}a^*_{\downarrow}]$, equal to the phase delay between the spin-up and spin-down ionisation pathways.

The amplitudes $a_{\uparrow}$, $a_{\downarrow}$, have been originally derived by U. Fano \cite{Fano1969}: $a_{\uparrow}=R_3$, $a_{\downarrow}=\frac{1}{3}(R_3+2R_1)$, where $R_{1,3}$ are the
radial transition matrix elements into the degenerate continuum states with total angular momentum $j=1/2$ and $j=3/2$.

To find the phase difference we need to find $\arg[a_{\uparrow}a^*_{\downarrow}]$:
\begin{multline}
a_{\uparrow}a^*_{\downarrow}=\left[R_3\left(\frac{1}{3}(R_3^{*}+2R_1^{*})\right)\right] \\= \frac{1}{3}\left(|R_3|^2+2|R_1||R_3|(\cos(\phi^{R}_3-\phi^{R}_1)+i\sin(\phi^{R}_3-\phi^{R}_1))\right),
\end{multline}
\begin{equation}
\Delta\phi_{SO}=\arg\left[R_3\left(\frac{1}{3}(R_3^{*}+2R_1^{*})\right)\right]=\arctan\frac{2|R_1||R_3|\sin(\phi^{R}_3-\phi^{R}_1)}{|R_3|^2+2|R_1||R_3|(\cos(\phi^{R}_3-\phi^{R}_1)}, \label{arg}
\end{equation}
Equation~\eref{arg} yields:
\begin{equation}
\tan\Delta\phi_{SO}=\frac{\sin(-\Delta\phi_{13})}{0.5|R_3|/|R_1|+\cos(\phi_{13})}, \label{Eq:Larmor1}
\end{equation}
where the phase difference $\Delta\phi_{13}$ is defined as $\Delta\phi_{13}=\phi^{R}_1-\phi^{R}_3$, the relative phase between $R_{1,3}(E)$.
 Their dependence on electron energy $E$ is very similar, up to a small off-set $\Delta E_{SO}$ due to the spin-orbit interaction in the ionisation channel: $R_3(E)=R_1(E-\Delta E_{SO})$ \cite{Fano1969}. The phases are shifted accordingly: $\phi^{R}_3(E)=\phi^{R}_1(E-\Delta E_{SO})$ \footnote{The photoionisation matrix elements we discuss here are complex, and we deal with their phases. The phase lag in a real-valued radial wave-function, corresponding to a given ionisation channel $j=3/2$ or $j=1/2$ discussed in \cite{Fano1969}, translates into the phase of the complex-valued photoionisation matrix element, leading to the phase difference  $\Delta\phi_{13}$ of the corresponding photoionisation matrix elements discussed here.} Using Taylor expansion, we find $\Delta\phi_{13}=\phi_1^{R}(E)-\phi_1^{R}(E-\Delta E_{SO})\simeq -\tau_{WS}\Delta E_{SO}$, where $\tau_{WS}=-d\phi_1^{R}/dE$ is the Wigner-Smith ionisation time \cite{Wigner1955,Smith1960,Maquet}. Thus, we have connected the angle of spin rotation during ionisation $\Delta\phi_{SO}$ to the Wigner-Smith ionisation time, calibrating our clock,
\begin{equation}
\tan\Delta\phi_{SO}=\frac{\sin(\tau_{WS}\Delta E_{SO})}{0.5|R_3|/|R_1|+\cos(\tau_{WS}\Delta E_{SO})}. \label{Eq:Larmor1a}
\end{equation}
The inhomogeneous character of the spin-orbit interaction makes the 
relationship between $\Delta\phi_{SO}$ and $\tau_{WS}$ nonlinear, introducing the extra term $0.5|R_3|/|R_1|$ in the denominator, but does not invalidate the clock. 

How does the spin-orbit interaction measure the ionisation time? 
The spin-orbit clock works like an interferometer, see Fig.~\ref{figure1}(b,c). The angle of rotation is given by the relative phase between the spin-up (parallel to the orbital momentum) and spin-down pathways. The spin-up pathway proceeds only via the $j=3/2$ continuum. The spin-down pathway is a double arm: it can proceed via both the $j=3/2$ and $j=1/2$ states. The spin-orbit interaction introduces the phase delay $\Delta\phi_{13}$ in the double arm. This small perturbation records the ionisation time, with Eq.~\eref{Eq:Larmor1a} connecting $\Delta\phi_{SO}$ and $\tau_{WS}$.

\section{Strong-field ionisation} \label{sec:iii}

We now turn to strong-field ionisation in intense IR fields--the regime of recent experiments\cite{Eckle2008,Shafir2012} aimed at measuring ionisation times and focus on the definition of the ionisation time in this regime.

\begin{figure}
\begin{center}
\includegraphics[width=12cm]{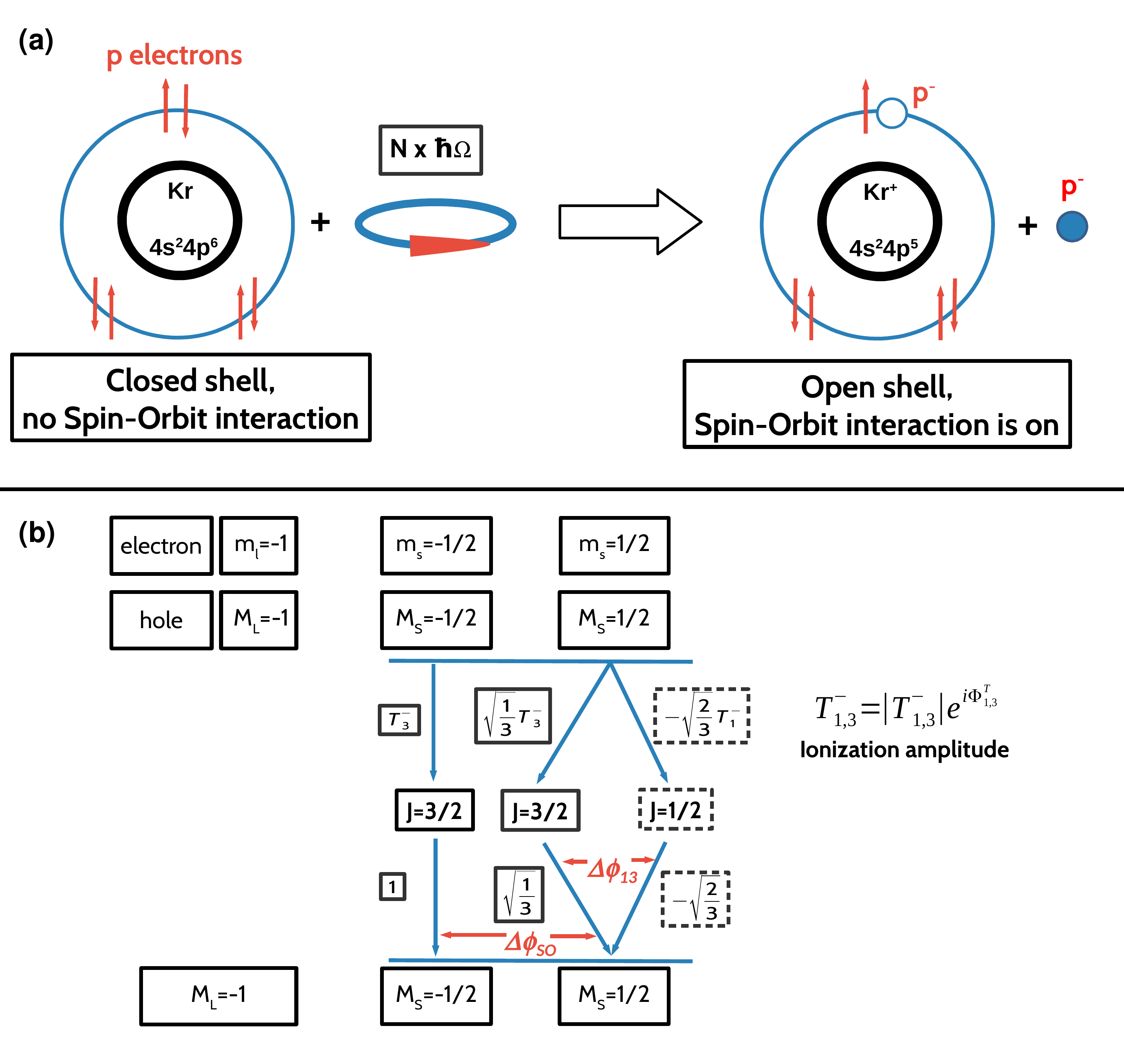}
\caption{Gedanken experiment for measuring ionisation time in strong-field ionisation. (a) Cartoon of an experiment, for a Kr atom: Intense circularly polarised field removes a $p^{-}$ electron from the valence 4$p^6$ shell via absorption of many photons, creating a rotating hole. (b) The spin-orbit clock operating on the hole as an interferometer, for an electron removed with $m_l=-1$ and the hole left with $M_L=-1$. As in Fig.~\ref{figure1}(c), the spin-up path is a double arm, since for $M_L=-1$ the spin-up pathway ($M_{S}=1/2$) can proceed via both $J=1/2$ and $J=3/2$ hole states. The ionisation amplitudes, up to the angular coefficients relating the orbital momentum $L$, spin $S$, and
the total angular momentum $J$, are $T_3^{-}$ (for $J=3/2$) and $T_1^{-}$ (for $J=1/2$). The relevant angular (Clebsch-Gordan) coefficients for each pathway are also indicated separately.} \label{figure2}
\end{center}
\end{figure}

To address this problem, consider ionisation of a Kr atom [Fig.~\ref{figure2}(a)]. The ground state of Kr$^+$ is spin-orbit split by the energy $\Delta E_{SO}=0.665$eV into the $P_{3/2}$ and $P_{1/2}$ states with total angular momentum $3/2, 1/2$. Both will be coherently populated
by ionisation \cite{Goulielmakis2010} in a few-cycle intense IR laser pulse, with coherence approaching 90\% for nearly single-cycle pulses
\cite{Goulielmakis2010}. The loss of coherence arises when photo-electron spectra, correlated to two different core states, do not overlap or overlap only partially. 

\begin{figure}
\begin{center}
\includegraphics[width=10cm]{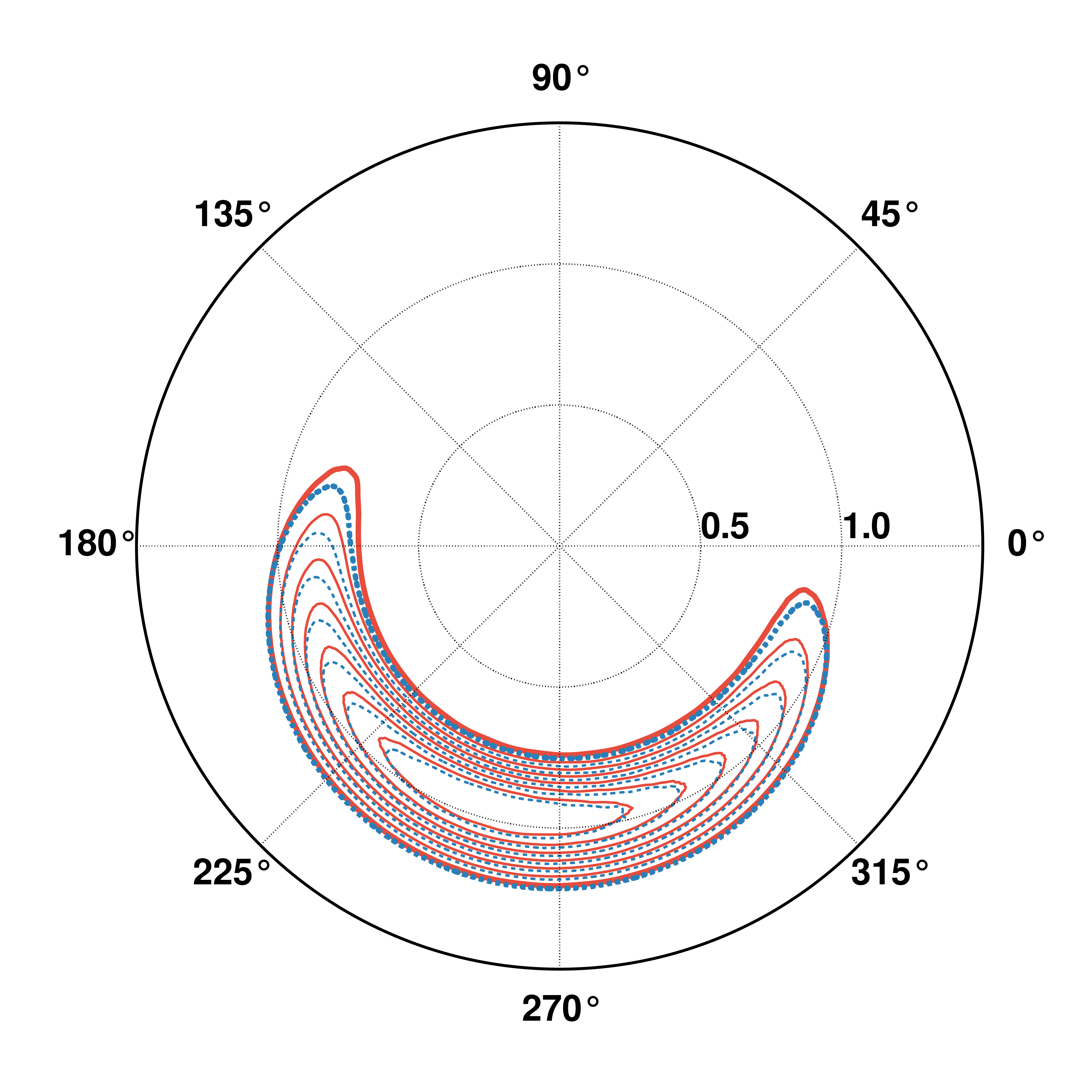}
\caption{Strong-field ionisation of Krypton by a a single-cycle, circularly polarised IR pulse. Angle and momentum-resolved photo-electron spectra calculated numerically for the two ionisation potentials corresponding to the two ground states of Kr$^+$, $P_{3/2}$ ($I_{p}=14.0$ eV) and $P_{1/2}$ ($I_{p}=14.67$ eV). Red solid contours correspond to P$_{3/2}$, blue dashed contours correspond to $P_{1/2}$. The inner-most contour corresponds to 0.9 level, other contours shown in steps of 0.1. The pulse had a vector-potential $\mathbf{A}_L(t) = -(F/\omega) \cos^4(\w t/4)(\cos\w t\,\mathbf{\hat{x}} + \sin\w t\,\mathbf{\hat{y}})$ with $F=0.05$ a.u. and $\omega=0.057$ a.u.
The radial coordinate gives the electron momentum in atomic units.} 
\label{figure3}
\end{center}
\end{figure}

Figure~\ref{figure3} shows nearly complete overlap of these spectra for the ultrashort circularly polarised pulse used in our ab-initio calculations, see \textcolor{blue}{\ref{app:i}} for the numerical details, confirming nearly 100\% coherence of the hole motion in this case. Note that the 
lack of 100\% coherence affects the amount of the 
coherently moving charge in the ion, but not the timing of its dynamics. 
Thus, even for coherences below 100\%, one is still able to use the rotation of the spin of the hole, triggered by the spin-orbit dynamics of the electron charge in the ion, as a clock. 

At the same time, the spin-orbit interaction in the ionisation channels becomes completely negligible in strong fields. The importance of this effect can be gauged using the Analytical $R$-matrix Approach (A$R$M) \cite{Torlina2012a,Torlina2012b,jivesh,lisa,attoclock}, which has been verified against ab-initio simulations in \cite{Torlina2014JPhysB,attoclock}. 
Application of ARM to calculating ionisation phases is described in \textcolor{blue}{\ref{app:ii}.} In the tunnelling limit, A$R$M yields the following expression for the relative phase between two degenerate continuum states associated with electron total momentum $j=l+1/2$ and $j^{'}=l-1/2$: $\xi_{SO}\sim\ 0.21\alpha^2 F^2/I_p^{5/2}\sim 2.3\times 10^{-7}$rad, which is completely negligible. Here $\alpha$ is the fine structure constant. For the estimates we used typical values of ionisation potential $I_p\simeq 0.5$ a.u. and the strength of the laser field $F=0.06$ a.u. (see \textcolor{blue}{\ref{app:ii}} for the details of the derivation).

The formal description of the spin-orbit interferometer in Fig.~\ref{figure2} is similar to the one-photon case in Fig.~\ref{figure1}. There is no spin-orbit interaction in the ground state of Kr: the filled valence $4p^{6}$-shell has equal number of $p^{-}$ and $p^{+}$ electrons `rotating' in opposite directions. Ionisation by a nearly single-cycle, right-circularly polarised IR pulse breaks the balance between the co-rotating and counter-rotating electrons: intense right-circularly polarised IR pulse preferentially removes the $p^{-}$ electron \cite{Barth2011,Herath2012,soarchive} and creates a $p^{-}$ hole. This starts the clock. The angle of rotation of the hole spin at a time-delay $\tau$ after the IR pulse is \textcolor{blue}{(see \ref{app:iii})}:
\begin{equation}
\tan\Delta\phi_{SO}=\frac{\sin(\Delta E_{SO}\tau -\Delta\phi_{13})}
{{0.5}|T^{-}_{3}|/|T^{-}_{1}|+\cos(\Delta E_{SO}\tau-\Delta\phi_{13})}.
\label{Eq:Larmor2}
\end{equation}
The dependence of the ionisation dynamics on the IR pulse intensity,
duration, shape, etc, is fully encoded in the matrix elements $T^{-}_{3}$ and $T^{-}_{1}$ [Fig.~\ref{figure2}(b)] describing strong-field ionisation amplitudes for the removal of $m_l=-1$ electron, leaving the hole in $p_{J,M_J}$ valence spin-orbitals with total angular momentum $J=3/2$ and $J=1/2$ correspondingly, for a given final electron momentum $\bf p$ at the detector. It corresponds to population of $P_{3/2}$ and $P_{1/2}$ states of Kr$^{+}$. The phases of $T_{3,1}^{-}$ are $\phi^{T}_{3}$ and $\phi^{T}_{1}$, $\Delta\phi_{13}=\phi^{T}_{1}-\phi^{T}_{3}$. For the $p^{+}$ electron ($m_l=1$) the expression is similar, except that $T_{1,3}^{+}$ are different. The tiny phase shift between spin-down and spin-up 
ionisation amplitudes correlated to the $P_{3/2}$ state of Kr$^+$ due to spin-orbit interaction in ionisation channels has the same estimate as above, $\xi_{SO}\sim\ 0.21\alpha^2 F^2/I_p^{5/2}\sim 2.3\times 10^{-7}$rad, and is also negligible.

The clock angles in Eqs.~\eref{Eq:Larmor1},~\eref{Eq:Larmor2} are virtually identical, except for the term $\Delta E_{SO}\tau$ describing the hole dynamics \cite{spin_currents} upon ionisation. The analogy in angle-time mapping in Eqs.~\eref{Eq:Larmor1} and \eref{Eq:Larmor2} allows us to establish the definition of strong-field ionisation time. Indeed, Eq.~\eref{Eq:Larmor1} calibrates the clock and establishes the mapping between the angle of spin rotation and the ionisation time. Eq.~\eref{Eq:Larmor2} contains the same mapping. Thus,
the time of hole formation is encoded in $\Delta\phi_{13}=\phi^{T}_{1}-\phi^{T}_{3}$, accumulated in the second (double) arm of the interferometer. We shall now analyse these phases to extract the strong-filed ionisation time. 

The phases $\phi^T_{1}, \phi^T_{3}$ encode the electron interaction with the potentials $U_{1,3}$ of the core states $P_{1/2}$ and $P_{3/2}$. These potentials have two contributions, $U_{1,3}=U^c+U^{d}_{1,3}$. Here $U^c$ is common for both states and is dominated by the long-range Coulomb potential, while
$U^{d}_{1,3}$ are different for the two core states, reflecting different spatial distributions of their electron densities, see \textcolor{blue}{\ref{app:iv}}. Thus, $\phi^T_{1,3}=\phi^c_{1,3}+\phi^{d}_{1,3}$.

For the same final kinetic momentum $\mathbf{p}$ of the continuum electron, in the strong-field ionisation regime, the difference between $\phi^c_{1}$ and $\phi^c_{3}$ comes from slightly different ionisation potentials into the $P_{1/2}$ and $P_{3/2}$ states: $\phi_1^c=\phi^c(I_p)$ and $\phi_3^c=\phi^c(I_p-\Delta E_{SO})$, see \textcolor{blue}{\ref{app:ii}}. Hence $\Delta\phi^c_{13}=\phi_1^c-\phi^c_3\simeq \Delta E_{SO}d\phi^c/d I_{p}$ and one should convert $\Delta\phi^c_{13}$ into time, dividing by the energy of spin-orbit splitting [note the `minus' sign in Eq.~\eref{Eq:Larmor2}]:
\begin{equation}
\tau_{SI}=-\frac{\Delta\phi_{13}^{c}}{\Delta E_{SO}}=-\frac{d \phi^{c}}{d I_p}. \label{Eq:Time}
\end{equation}
Equation~\eref{Eq:Time} defines ionisation time in the strong-field regime. 

The second part of the relative phase, $\Delta\phi^{d}_{13}=\phi_1^d-\phi^d_3$, results from the different core potentials for the $P_{3/2}$ and $P_{1/2}$ core states, e.g. due to the different angular structure of the electron density. It does not depend on $\Delta E_{SO}$, i.e. the period of the clock, and hence can not be converted into the time-delay in the formation of the hole.

We note that the derivation presented here is not applicable in the weak-field regime.
Firstly, the neglected spin-orbit interaction in ionisation channel may become important in weak fields. However, it is a rather standard approximation to ignore spin-orbit interaction in the ionisation channel compared to the spin-orbit interaction for the core electrons (see e.g. \cite{santra}), since the core electrons are plenty and stay near the core, where the spin-orbit interaction is strong, while the sole continuum electron leaves the core region.
Secondly, the explicit dependence of phases $\phi_{1,3}$ on ionisation potential, used in deriving Eq.~\eref{Eq:Time}, arises naturally only in the strong-field regime, see \textcolor{blue}{\ref{app:ii}}. Therefore, the expression Eq.~\eref{Eq:Time} may not hold in the weak field regime.

\section{Reading spin-orbit Larmor clock in strong-field regime} \label{sec:iv}

In the Gedanken experiment described in Section~\ref{sec:iii}, Eq.~\eref{Eq:Larmor2} is sufficient to introduce the strong-field ionisation time $\tau_{SI}$ by comparing it to Eq.~\eref{Eq:Larmor1a} which has calibrated the clock. In contrast to the one-photon case, where the clock stops as soon as the liberated electron leaves the range of the spin-orbit interaction potential, the discussion in Section~\ref{sec:iii} does not involve stopping the clock. Indeed, the clock operates on the hole states $J=3/2$ and $J=1/2$. The hole spin periodically rotates after ionisation is completed. Thus, the clock continues to count time after it has recorded the rotation angle related to the ionisation time. To stop the clock and read the information out we can apply the second pulse. It allows us to get direct access to the phase $\Delta\phi_{13}$, which records time in the spin-orbit interferometer shown in Fig.~\ref{figure2}(b).

\begin{figure}
\begin{center}
\includegraphics[width=12cm]{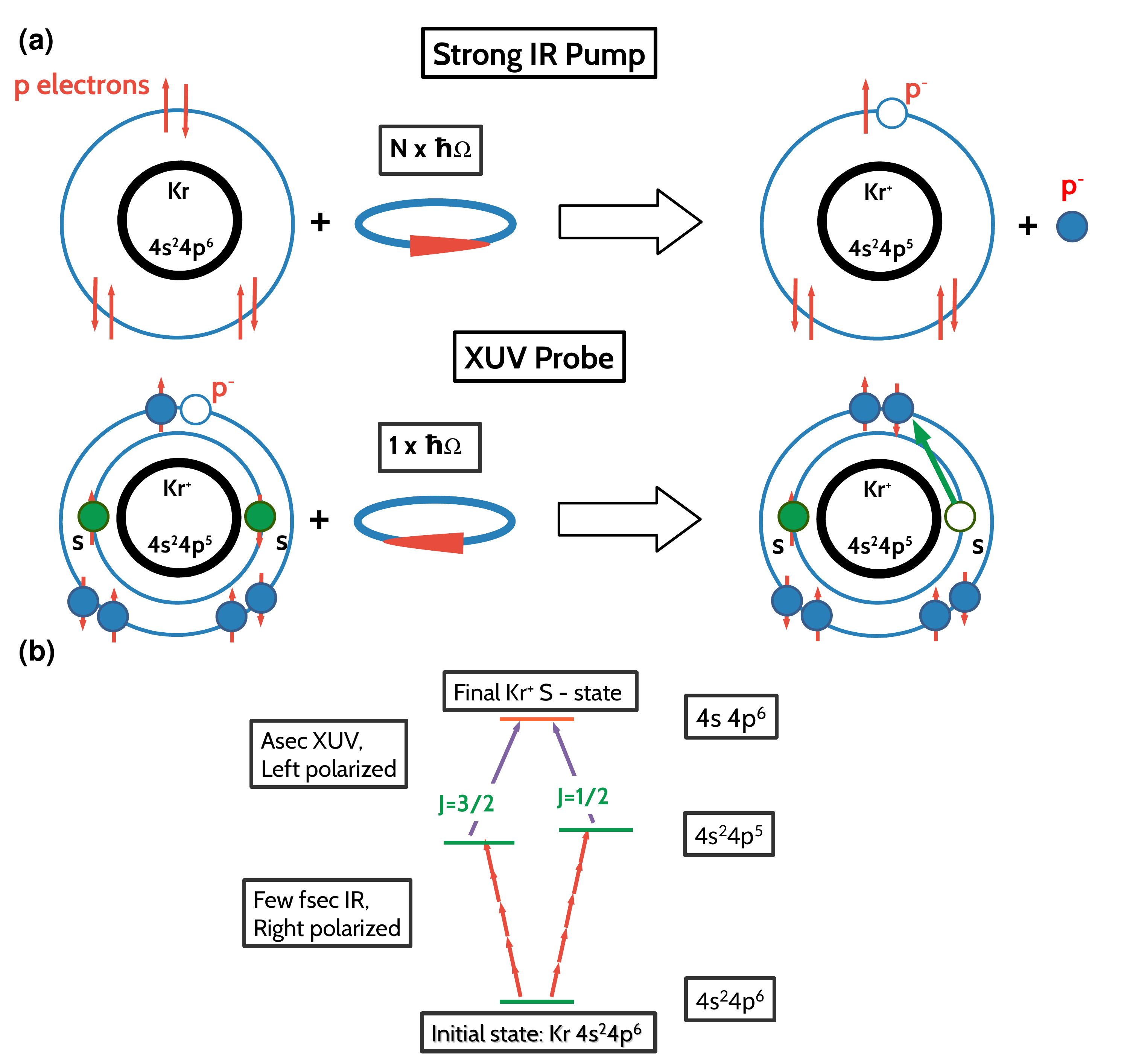}
\caption{Schematic of a laboratory pump-probe experiment implementing the spin-orbit Larmor clock for strong-field ionisation, for Kr atom. (a) Cartoon of the experiment. Multiphoton ionisation with a strong, right-circularly polarised infrared pump pulse creates a $p^{-}$-hole 
and starts the clock. Attosecond extreme ultraviolet probe fills the $p^{-}$-hole by promoting an electron from the inner $s$-shell. This transition stops the spin-orbit clock, since spin-orbit interaction for $s$-states is absent. (b) Analysis of the experiment as a two-path interferometer. Two pathways via the $P_{1/2}$ and $P_{3/2}$ states of the ion interfere in the final $S$-state of the Kr$^+$. }
\label{figure4}
\end{center}
\end{figure}

Here is how it works. Consider the pump-probe experimental scheme shown in Fig.~\ref{figure4}(a,b). The pump, which starts the clock, is a nearly single cycle right circularly polarised IR pulse. The probe, which stops the clock, is a left circularly polarised attosecond XUV pulse. It comes with an attosecond-controlled delay $\tau$ and fills the hole in $p^{-}$ orbital by exciting an electron from a deeper $s$-orbital (see Fig.4(a)). More generally, the probe pulse promotes the core into an excited $S$-state, where the spin-orbit splitting is absent, e.g. $4s4p^6$ or any other suitable state. Broad bandwidth of the attosecond probe pulse couples both $P_{3/2}$ and $P_{1/2}$ to the same final $S$-state, as in \cite{Goulielmakis2010} (see Fig. 4(b)). As opposed to the Gedanken experiment above, in laboratory experiments the initial spin-up and spin-down components of the ground state are incoherent, and the single arm of the interferometer yields background for the interference in the double arm in Fig.~\ref{figure4}(b). 
Left-polarised probe ensures that the final $S$-state can be reached only if the electron missing in the Kr core after ionisation is the $p^-$ electron.

The population $w$ of the final state is \textcolor{blue}{(see \ref{app:v})}:
\begin{equation}
\label{Eq:Transition}
 w=|A_1|^2+|A_3|^2+2|A_1||A_3|\cos(\Delta E_{SO}\tau-\Delta\phi_{13}(\p)) +|\tilde A_3|^2.
\end{equation}
Here, $A_1=2T^{-}_{1}(\p)d_{1/2}F_{\omega}(\Omega_1)\sqrt{2/27}$ and $A_3=T^{-}_{3}(\p)d_{3/2}F_{\omega}(\Omega_3)\sqrt{2/27}$ are the transition amplitudes for the two interfering pathways corresponding to the removal of the spin-down $p^{-}$ electron. In addition to the multi-photon ionisation matrix elements $T^{-}_{1,3}(\p)$
they include the real-valued radial transition matrix elements $d_{1/2}$ and $d_{3/2}$ between the $P_{1/2}, P_{3/2}$ and the final $S$-state of Kr$^+$, and the spectral amplitudes of the attosecond pulse, $F_{\omega}(...)$, at the excitation energies $\Omega_{3,1}$ from the $P_{3/2, 1/2}$ states to the final $S$-state. The background $|\tilde A_3|^2$, $\tilde A_3=\tilde T^{-}_{3}(\p)d_{3/2}F_{\omega}(\Omega_3)\sqrt{2/27}$,
corresponds to the removal of the \textit{spin-up} $p^-$ electron \textcolor{blue}{(See \ref{app:v}} for details of the derivation). Modulation of $w$ versus the pump-probe delay $\tau$ yields the phase $\Delta\phi_{13}(\p)$ between $A_1$ and $A_3$. It can be measured, e.g. using attosecond transient absorption \cite{Goulielmakis2010}. The sensitivity of  the phase $\Delta\phi_{13}(\p)$
to the final momentum $\p$, and the errors that it can introduce into the transient absorption 
measurement of the ionization time, are discussed below.

The phase $\Delta\phi_{13}$ includes two contributions: (i) the relative phase due to the same core potential in both ionisation pathways, $\Delta\phi^c_{13}$, which can be translated into time-delay and (ii) the phase $\Delta\phi^d_{13}$, related to the different electron-core potentials in the two ionisation channels. This phase reflects correlation between the electron and the core and  can not be translated into time. If an experiment does not distinguish between these two contributions to $\Delta\phi_{13}$, the phase $\Delta\phi^d_{13}$ related to the electron-hole correlation  
will look like a time shift.

\begin{figure}
\begin{center}
\includegraphics[width=13cm]{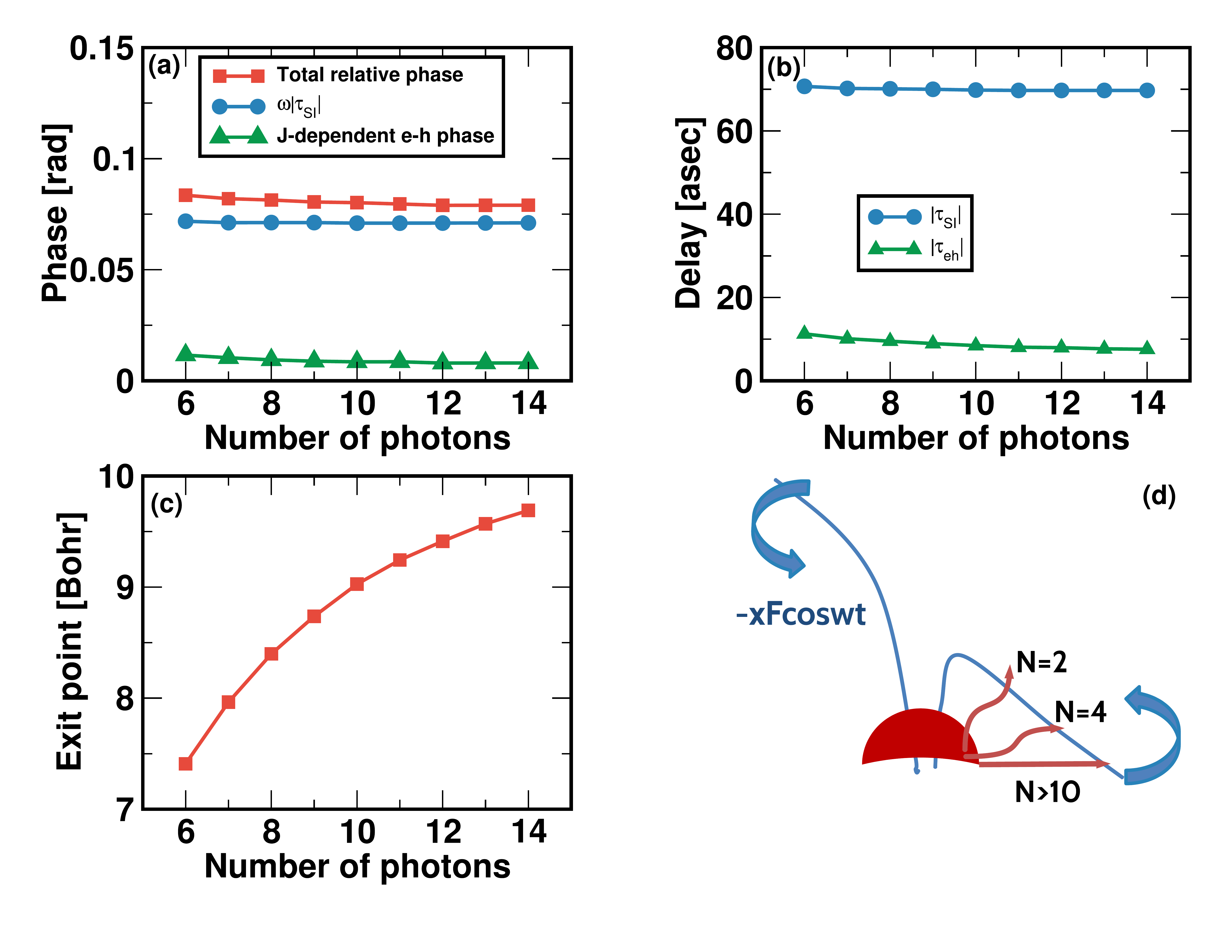}
\caption{Analysis of time delays in strong-field ionisation.
(a) Calculated phase $\Delta \phi_{13}$ for the pump-probe experiment, as a function of the minimum number of photons required for ionisation, $N=I_p/\omega$, $I_p$ is ionisation potential, $\omega$ is laser frequency. The calculations were done for a Kr atom and the circular field intensity $2.5\times 10^{14}$W/cm$^2$. Blue circles show phase associated with the actual time-delay. Green triangles show the phase that does not correspond to time-delays but is a leftover from the electron-hole correlation. Total phase is shown as red squares. 
(b) Real (blue circles) and 'apparent' (green triangles) ionisation delays as a function of the number of photons required for ionisation, $N$. (c,d) Physical picture underlying the results: $N$-dependence of the electron exit position from the potential well (see \ref{app:ii}) (c) and the cartoon of the ionisation process (d).} \label{figure5}
\end{center}
\end{figure}

Figure~\ref{figure5}(a) shows how the total phase (red squares), which can be measured by transient absorption, and its two separate parts $\Delta \phi^c_{13}$ (blue circles) and $\Delta\phi^d_{13}$ (green triangles), depend on the laser wavelength, i.e. the minimum number of photons 
$N=I_p/\omega$ required to reach the ionisation threshold, for fixed laser intensity. Figure~\ref{figure5}(b) shows $\tau_{SI}$ (blue circles) and apparent time delays $\tau_{eh}=-\Delta \phi^{d}_{13}/\Delta E_{SO}$ (green triangles). The apparent delay $\tau_{eh}$ is not negligible for $P_{3/2}$ and $P_{1/2}$ states of Kr$^{+}$.
 
To obtain results in Fig.~\ref{figure5}(a,b), we have calculated the phases accumulated due to the Coulomb potential and the short range components of the core potential for the two ionisation channels, corresponding to the ionic states $P_{3/2}$ and $P_{1/2}$. Note that
the short-range potentials in these two channels are different, see \textcolor{blue}{\ref{app:iv}.} To obtain time-delays, we have divided the relative phases by the difference in the ionisation potentials, $\Delta E_{SO}$. The phases were calculated using the A$R$M method \cite{Torlina2012a,Torlina2012b,jivesh,lisa,attoclock}, for the characteristic momentum of the photo-electron distribution $p_0=A_0 \sinh(\omega \tau_T)/(\omega \tau_T)$, where $A_0$ is the amplitude of the field vector potential and $\tau_T\equiv {\rm Im}[t_s(p_0)]$ is the so-called `Keldysh tunnelling time', the imaginary part of the saddle point $t_s(p_0)$, see \textcolor{blue}{\ref{app:iii}.} For this momentum, which is very close to the peak of the distribution for the short laser pulse, the ionisation phases
have simple analytical expressions in the tunnelling limit:
\begin{gather}
\Delta\phi^c_{13}\simeq -\Delta E_{SO}/I_p^{3/2}, \\
\Delta\phi^d_{13}\simeq -0.4 F^2/I_p^{5/2}. \label{eq:TunnellingPhase}
\end{gather}
Note that $\Delta\phi^c_{13}$ is proportional to $\Delta E_{SO}$ and therefore leads to proper time-delay, while the phase $\Delta\phi^d_{13}$ accumulated due to the different short-range potentials does not scale with $\Delta E_{SO}$ and cannot be translated into proper time. \textsf{Thus, every time is phase, but not every phase is time}. 

\begin{figure}
\begin{center}
\includegraphics[width=12cm]{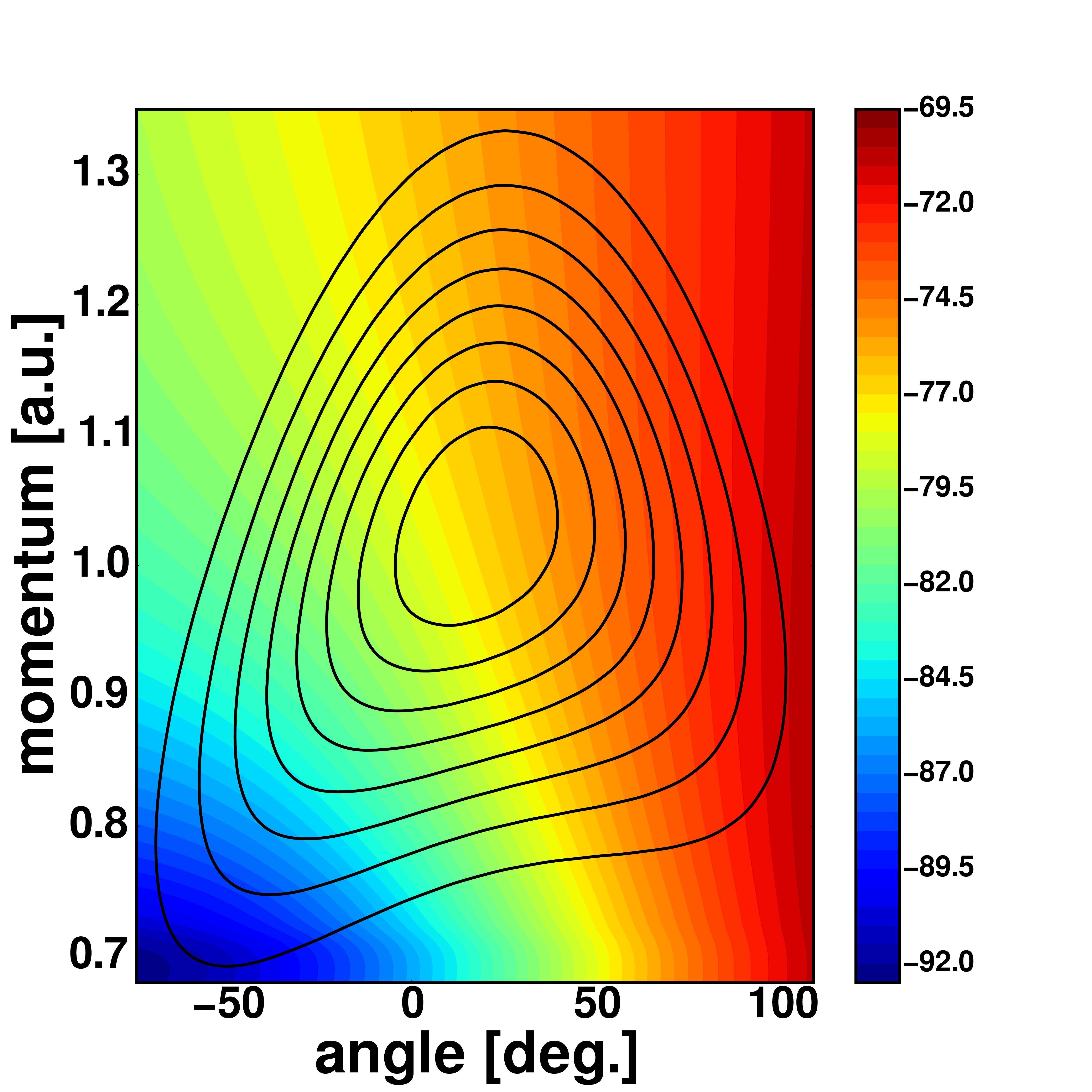}
\caption{Dependence of ionisation delays $\tau_{SI}$ on the electron momentum at the detector, for Hydrogen atom. The inner-most contour in the electron spectrum corresponds to 0.9 level, other contours are shown in steps of 0.1. The color bar shows $\tau_{SI}$ in attoseconds. 
The pulse had a vector-potential $\mathbf{A}_L(t) = -(F/\omega) \cos^4(\w t/4)(\cos(\w t)\,\mathbf{\hat{x}} + \sin(\w t)\,\mathbf{\hat{y}})$ with $F=0.05$ a.u. and $\omega=0.0465$ a.u. Results are obtained using the A$R$M theory.} \label{figure6}
\end{center}
\end{figure}

Since transient absorption experiments do not detect the final energy (or the momentum) of the electron, we have also checked that the phases and the resulting times are only very weakly sensitive to the final electron momenta within the region surrounding the peak of the photo-electron signal. This analysis is presented in Fig.~\ref{figure6}, where the ionisation time-delays are overlayed with the electron spectrum generated by the single-cycle pump pulse with the vector-potential $\mathbf{A}_L(t) = -A_{0}\cos^4(\w t/4)(\cos(\w t)\,\mathbf{\hat{x}} + \sin(\w t)\,\mathbf{\hat{y}})$, 
with $A_0=F/\omega$, $F=0.05$ a.u. and $\omega=0.0465$ a.u.

The difference in ionisation times within the full width at the half-maximum of the distribution is $\pm 5$ asec. This number provides an estimate for possible errors in transient absorption measurements of ionisation delays caused by averaging over the photo-electron distribution. Note that such measurements will also inevitably include the apparent delays $\tau_{eh}=-\Delta\phi^d_{13}/\Delta E_{SO}$ associated with the phase $\Delta\phi^d_{13}$. For the specific example shown in Fig.~\ref{figure5}(b) $\tau_{eh}\sim 10$ asec. Importantly, in the tunnelling limit $\tau_{SI}$ is intensity-independent while the apparent delay $\tau_{eh}$ is proportional to the laser intensity, see Eq.~\eref{eq:TunnellingPhase}. This factor might be used to separate these two contributions.

Results presented in Fig.~\ref{figure5}(b) show that, as we increase the laser wavelength $\lambda$ and hence the number of photons $N=I_p/\omega$ required for ionisation, the ionisation time in Fig.~\ref{figure5}(b) decreases. This dependence has simple explanation. As $\lambda$ decreases, the laser frequency $\omega$ increases, ionisation becomes less adiabatic and the electron splashes out of the potential well closer to the core, see Fig.~\ref{figure5}(c,d). From there, it runs to the detector, accumulating the phase and consequently the time-delay $\tau_{SI}$. The closer the electron is launched, the larger is the accumulated phase. Note that no delay is accumulated under the barrier, see Section~\ref{sec:vi} for details.

\section{Attoclock measurements of strong-field ionisation delay} \label{sec:v}

The spin-orbit Larmor clock has offered us a general procedure for defining ionisation times in both one-photon and strong field ionisation regimes. Using the same general procedure we have arrived at two different expressions in the weak-field one-photon ionisation regime and in the strong-field regime. In the weak field regime we found the Wigner-Smith ionisation time. In the strong-field regime we found an expression which agrees with the result of a completely different derivation described in \cite{attoclock}. Importantly, while we have derived ionisation times using spin-orbit interaction, our results do not depend on it. Therefore, the detection of the strong-field ionisation time does not have to rely on the spin-orbit interaction.

Consider, for example, the so-called attoclock setup \cite{Eckle2008,Pfeiffer2012}, which measures angle-resolved electron spectra produced in nearly circular, few-femtosecond IR pulses. Such pulses send electrons released at different instants of time in different directions, providing the link between the direction of electron velocity at the detector and the time of its release. Nearly single-cycle pulse creates preferred direction of electron escape, from which the ionisation delay can be reconstructed \cite{Eckle2008,Pfeiffer2012,attoclock}. The angle $\phi_{\text{max}}$ at which the majority of electrons are detected, relative to the detection angle expected in the absence of the core potential, is called the off-set angle. We now show that $\phi_{\text{max}}$ can measure the ionisation delay $\tau_{SI}$ derived above, provided that effects leading to transient population trapping of released electrons and "negative" ionisation times \cite{attoclock} are negligible. In particular, such regime can be achieved in the long wave-length limit (but is not limited to it).

\begin{figure}
\begin{center}
\includegraphics[width=13cm]{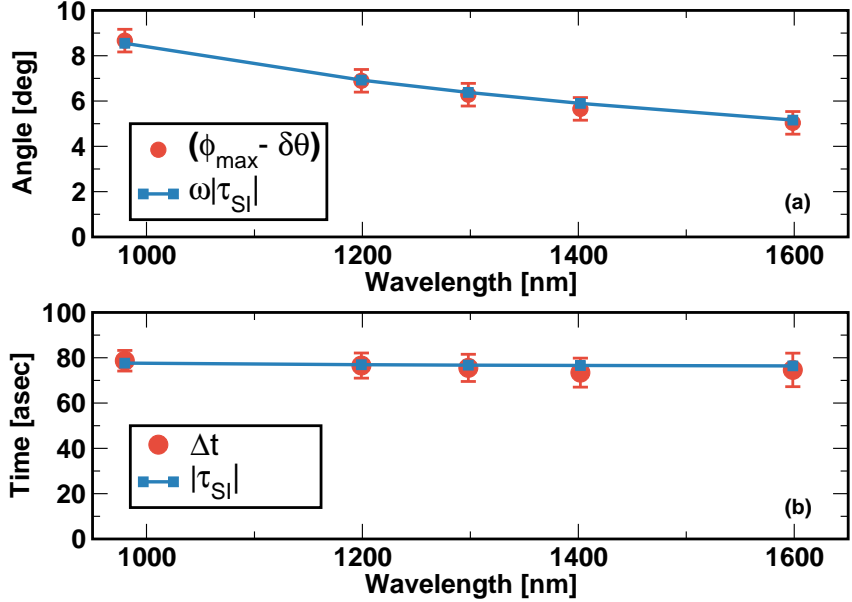}
\caption{Attoclock measurements of strong-field ionisation delay. (a) Red circles show numerically calculated envelope-free offset angle $\phi_{max}-\delta \theta$. The small correction $\delta \theta$ to the off-set angle $\phi_{max}$ is introduced by the rapidly changing pulse envelope of the nearly single-cycle laser pulse \cite{attoclock} and is subtracted from $\phi_{max}$ to present envelope-free results for the offset angle. The blue squares connected by the blue line show $\omega |\tau_{SI}|$, the offset angle corresponding to time-delay $|\tau_{SI}|$, $\omega$ is the laser frequency. (b) Red circles show the ionisation delay $\Delta t=(\phi_{max}-\delta \theta)/\omega$ extracted from the ab-initio photoelectron spectra. The blue squares connected by the blue line show $\tau_{SI}$. All calculations were done for a hydrogen atom and the circular field intensity $1.75\times 10^{14}$W/cm$^2$.} \label{figure7}
\end{center}
\end{figure}

To this end, we consider the benchmark system--the hydrogen atom, where  fully ab-initio simulation of ionisation dynamics in the strong circularly polarised IR field is possible. We solve the time-dependent Scr{\"o}dinger equation numerically exactly and use results of the numerical experiment to find $\phi_{\text{max}}$. Details of the calculation are described in \textcolor{blue}{\ref{app:i}.} Red circles in Fig.~\ref{figure7} show the ionisation delay $\Delta t=(\phi_{\text{max}}-\delta \theta)/\omega$ extracted from the ab-initio photoelectron spectra, where $\omega$ is the laser frequency and the small correction $\delta \theta$ to the off-set angle $\phi_{max}$ is introduced by the rapidly changing pulse envelope of the nearly single-cycle laser pulse we have used \cite{attoclock}. The blue curve with squares, which shows $\tau_{SI}$ Eq.~\eref{Eq:Time}, lies on top of the ab-initio results. To calculate $\tau_{SI}$ analytically, we have used the A$R$M theory \cite{attoclock}.

We stress that the definition of $\tau_{SI}$ is not restricted to the 
long-wavelengths limit shown in Fig.~\ref{figure7}. It is only the ability of the attoclock set-up to measure exclusively this time delay, without additional contributions associated with 
transient population trapping in Rydberg states leading to negative 
ionisation times \cite{attoclock}, that has restricted our consideration to the  wavelength regime shown in Fig.~\ref{figure7}. Nevertheless, it is important to demonstrate at least one example, where the time delay $\tau_{SI}$ derived from the idea of the spin-orbit Larmor clock can be experimentally or numerically detected.

\section{Strong-field ionisation delay and tunnelling delay} \label{sec:vi}

\begin{figure}
\begin{center}
\includegraphics[width=7cm]{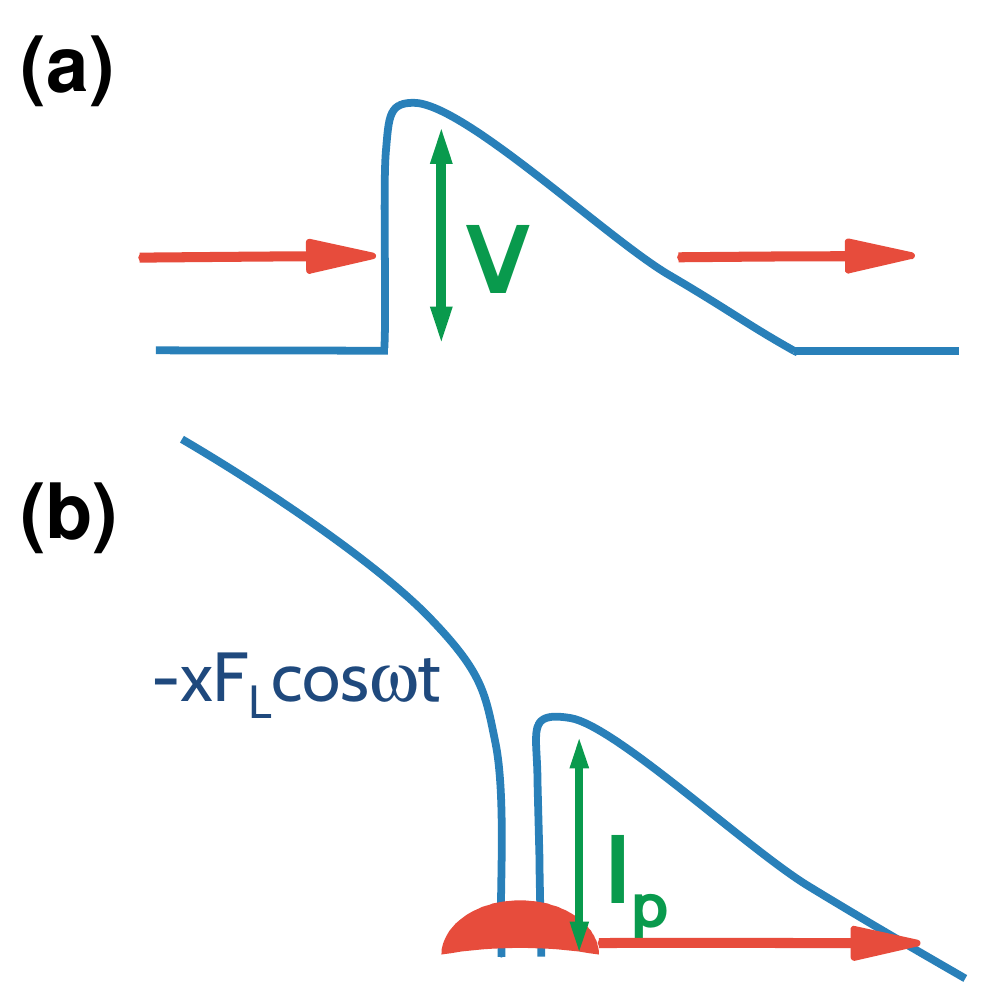}
\caption{Cartoon illustrating the analogy and the difference between
(a) the standard barrier penetration problem, and (b) optical tunnelling through the barrier created by the laser field and the core potential in strong-field ionisation.} \label{figure8}
\end{center}
\end{figure}

Strong-field ionisation is often viewed as tunnelling through the barrier created by the binding potential and the laser electric field. While our analysis has never relied on the tunnelling picture, our definition is consistent with the Larmor time $-\partial \phi/\partial V$ for tunnelling through a static barrier of height $V$ \cite{Landauer1994,Hauge}, equal to $I_p$ in our case (see Fig.~\ref{figure8}(a,b)), $\phi$ is the phase of electron wave-function. However, Fig.~\ref{figure8}(a,b) emphasises the difference in the two processes, which is in the boundary or initial conditions for the tunnelling dynamics. In Fig.~\ref{figure8}(a), the electron current is incident on the barrier and it can lead to the appearance of tunnelling delays, i.e. phase and time delays accumulated during the motion under the barrier. In Fig.~\ref{figure8}(b) the tunnelling starts from the real-valued wave-function of the bound state. It is a plausible assumption that in this case and for the low-frequency laser field, the polarized bound state carries negligible current incident on the barrier, and that tunnelling would occur from the tail of the initial wave-function already present under the barrier. Since the initial wave-function is real-valued in the barrier region, the phase in Eq.~\eref{Eq:Larmor1} may get no contribution from the tunnelling region, leading to no delay associated with the under-barrier part of the electron motion. Indeed, the analytical calculation of the phase in Eq.~\eref{Eq:Larmor1} yields no contribution from the under-barrier region, at least in the regime of Fig.~\ref{figure7}.

As it follows from excellent agreement between analytical and numerical results in Fig.~\ref{figure7}(a,b), the analytical calculation of the phase is accurate, and optical tunnelling is not associated with time delay. The delay $\tau_{SI}$ is only due to electron interaction with long-range core (Coulomb) potential and is explicitly accumulated after the exit from the barrier.

\section{Conclusions} \label{sec:conc}

We have illustrated the concept and the meaning of time delays in strong field ionisation. In one electron systems, these delays are related to electron interaction with the nucleus. In the tunnelling limit, comparison of numerical and analytical results unambiguously demonstrates the absence of tunnelling delays. Non-equilibrium charge dynamics excited in a many electron atom or a molecule by the laser field and the electron-electron correlations \cite{Torlina2012b,Walters2010} could lead to additional phase $\delta \phi$ \cite{Mairesse2010} and additional delays 
$\delta \tau_{SI}=-d \delta \phi/dI_{p}$ contributing to $\tau_{SI}$. Our work shows why and how ionisation delays provide a window into such dynamics in complex systems.

Production of a coherent superposition of many ionic states and hence of coherent hole dynamics is the key aspect of interaction with ultra-short light pulses. Any pump-probe experiment resolving these dynamics aims to find phases between the coherently populated states. As a result of electron-core correlations, not all phases are mapped into time: the formation of the hole wavepacket is characterised not only by the overall time-delays, but also by additional phases accumulated during the ionisation process due to the
different core potentials for the different final states of the ion. 

What do these phases mean? 
Given that the electron wavepackets correlated to different core states overlap at the detector, the hole presents a coherent wavepacket characterised by the relative phases of its different spectral components. Analysis of spectral phase is common in characterisation of ultrashort pulses in optics. Linear spectral phase records the arrival time, while non-linear phase is associated with pulse dispersion. Such dispersion is the closest analogue of the phase shifts related to electron-core correlations.

\ack
We thank Alfred Maquet and Armin Scrinzi for stimulating discussions.
J.K., O. S. and M. I. acknowledge support of the EU Marie Curie ITN network CORINF, Grant Agreement No. 264951. F. M. and O. S. acknowledge support of the DFG project SM 292/3-1, M. I. acknowledge support of the EPSRC Programme Grant EP/I032517/1, and the United States Air Force Office of Scientific Research program No. FA9550-12-1-0482, O.S., L. T. and J.K. acknowledge support of the DFG grant SM 292/2-3.


\appendix

\section{Ab-initio calculations} \label{app:i}

The numerical procedure and the code are described in detail in \cite{muller,attoclock}. The method has been monitored for convergence
by changing the maximum angular momentum up to $L_{\text{max}}=300$, the radial grid size was increased up to $r_{\text{max}} = 2500$ a.u., and by varying the step size of the radial grid $\delta r$ down to 0.05 a.u. In the presented calculations, the step size of the radial grid was $\delta r = 0.15$ a.u., the 
time-step was $\delta t = 0.04$ a.u., the box size was 1500 a.u., and $L_{\text{max}}=150$.

For Hydrogen atom, the spectrum was obtained by projection on the exact field-free continuum states of the $H$-atom after the end of the laser pulse. The photoelectron spectra include the volume element $\propto p^2$, both in numerical and analytical calculations. The volume element shifts the position of the peak of the distribution and thus affects the off-set angle, however, in the exact same way for numerical and analytical spectra. In these both numerical and analytical calculations we define the laser field $\F_L(t)=-{\partial\mathbf{A}_L(t)}/{\partial t}$ via the vector-potential $\mathbf{A}_{L}(t)$:
\begin{equation}
\mathbf{A}_L(t) = -A_0 f(t)(\cos(\w t)\,\mathbf{\hat{x}} + \sin(\w t)\,\mathbf{\hat{y}}), \label{pulse}
\end{equation}
where $f(t)$ is the pulse envelope and $\w$ is the carrier frequency,
\begin{equation}
f(t)=\cos^4(\w t/4). \label{envelope}
\end{equation}

For Kr atom, the calculations 
have been performed using the effective one-electron model potential
\begin{equation}
U_{Kr}(r) = \frac{1 + (36 - 1)*\exp(-\eta\,r)}{r}+U_0,
\end{equation}
based on the DFT potential used by D. Bauer and co-workers \cite{Bauer}. 
We follow the recipe described  in Ref.\cite{muller1}, using 
the additional tuning potential $U_0$ which is added only at the first radial 
grid point $r_1=0.5$ 
(the radial grid step was $\Delta r=0.05$ a.u.) and is equal to zero
everywhere else. 
The parameter $\eta=2.64343586965$ a.u.  
has been adjusted to yield the correct ionisation 
potential of Kr for the lowest $J=3/2$ ionic state,
with additional fine-tuning achieved by setting $U_0= 0.0249$a.u., giving
$I_p = -0.5145022731$ a.u. For the  $J=1/2$ core state the tuning 
potential was adjusted to $U_0=22.7629$ a.u., yielding
$I_p = -0.5389895221$ a.u. 

The photoelectron spectrum was calculated by propagating the wavefunction for sufficiently long time after the end of the laser pulse (typically 2 cycles, the convergence has been monitored up to 10 cycles), 
then applying a spatial mask to filter out the central part of the wavepacket within 100 Bohr from the origin, and performing the Fourier transform of the remaining part of the wavepacket. We have independently validated this procedure using the Hydrogen atom, where it has been calibrated against the projection of the wavefunction on the exact scattering continuum states for Hydrogen. The mask radius was chosen based on this calibration in Hydrogen.

\section{Calculation of the phase accumulated due to interactions in ionisation channels} \label{app:ii}

\subsection{Definition of the strong-field ionisation phase accumulated due to interactions in ionisation channels} \label{subapp:phiIC}

To evaluate the relative phase between the two ionisation channels in Kr,
we use the $R$-matrix based method (A$R$M) \cite{Torlina2012a,Torlina2012b} generalised for the case of circularly polarised fields \cite{jivesh,lisa,attoclock}.

The A$R$M method allows one to obtain an analytical expression for the total phase accumulated in each ionisation channel:
\begin{equation}
\phi_J(\p,t_s(\p,I_p)) = \int_{t_s - i\kappa^{-2}}^{T} dt\,U_{J}\left(\int_{t_s}^{t}d\zeta\,\mathbf{v}(\zeta)\right), 
\label{eq:WC}
\end{equation}
where $U_J(\r)$ is the potential defining the interaction,
$\kappa=\sqrt{2I_p}$, $I_p$ is the ionisation potential, $\mathbf{v}(t)=\mathbf{p}+\mathbf{A}(t)$, $\mathbf{A}(t)$ is vector-potential of the laser field, $T\rightarrow\infty$ is the observation time and $\mathbf{p}$ is the electron final momentum at the observation time.

The time $t_s(\p,I_p)$ (see \cite{jivesh, attoclock}) is the complex-valued solution of the saddle point equations for the ionisation in circularly polarised field:
\begin{equation}
\frac{\partial S_{\text{V}}(T,\p,t_s)}{\partial t_s} = I_p,\label{saddle}
\end{equation}
where $S_{V}(T,\p)$ is the Volkov phase accumulated by the electron in the laser field only: 
\begin{equation}
S_{V}(T,\p,t_s) = \frac{1}{2}\int_{t_s}^{T}dt\,[\p+\A(t)]^2. \label{Sv}
\end{equation}
The coordinate of exit presented in Fig.~\ref{figure5}(c) of the main text is:
\begin{equation}
\mathbf{r}_0 = \int_{t_s}^{\text{Re}[t_s]}d\zeta\,\mathbf{v}(\zeta).
\end{equation}
Since $t_s(\p,I_p)$ depends on $I_p$, the phase $\phi_J(\p,t_s(\p,I_p))$ also depends on $I_p$. The phase difference in the two channels is accumulated due to the different $I_p$'s: the difference in ionisation potentials leads to slightly different $t_s$ and thus slightly different trajectories in the two channels.
These trajectories are the arguments of $U_J$ in Eq.(\ref{eq:WC}). The common part of the phase is accumulated due to the Coulomb potential. The channel-specific part is accumulated due to the channel-specific core potential discussed in \textcolor{blue}{\ref{app:iv}}. The phase accumulated due to spin-orbit interaction in the ionisation channel is negligible and is estimated below \textcolor{blue}{in \ref{subapp:phiSO}}.

\subsection{The phase accumulated due to spin-orbit interaction in ionisation channel} \label{subapp:phiSO}

We estimate the relative phase between the two ionisation channels corresponding to spin-up and spin-down ionisation pathways, with orbital momentum $l$ and two values of electron total momentum: $j=l+1/2$, and $j'=l-1/2$. We use the spin-orbit interaction potential:
\begin{equation}
V_{SO}(r)=-\frac{j(j+1)-l(l+1)-s(s+1)}{4c^2r^3},
\end{equation}
where $c = 1/\alpha \approx 137$, in atomic units, $s = 1/2$ is electron spin ($\alpha$ being the fine-structure constant). The phase difference $\xi_{SO}$ is expressed via the difference between the potentials corresponding to $j$ and $j'$: 
\begin{equation}
\Delta V_{SO}(r)=-\frac{l+1/2}{2c^{2}r^{3}},
\end{equation}
We now calculate the phase difference using Eq.~\eref{eq:WC} connecting the phase to the potential. Substituting the electron trajectory in the tunnelling limit $r=r_0+Ft^2/2$, where $r_0=I_p/F$, $F$ is the field strength, we obtain the following integral:
\begin{equation}
\xi_{SO}=-\frac{(l+1/2)}{2c^2r_0^3}\int_0^{\infty} \frac{dt}{(1+\frac{Ft^2}{2r_0})^3},
\end{equation}
where $l$, the electron angular momentum along the trajectory, remains constant in the pure tunnelling limit. 
Evaluating the integral:
\begin{equation}
\int_0^{\infty} \frac{dt}{(1+\frac{Ft^2}{2r_0})^3}=\frac{\sqrt{2r_0F}}{F}\int_0^{\infty} \frac{dx}{(1+x^2)^3}=0.59\frac{\sqrt{2I_p}}{F},
\end{equation}
we obtain:
\begin{equation}
\xi_{SO}=-\frac{(l+1/2)}{2c^2r_0^3}0.59\frac{\sqrt{2I_p}}{F}=-0.42\frac{(l+1/2)}{c^2}\frac{F^2}{I_p^{5/2}}.
\end{equation}	
Note that in general the angular momentum of the electron $l$ is changing with time and should be included in the integrand. However, the integral is accumulated in the vicinity of the core	and therefore for  estimates in the tunnelling limit we can use the value of angular momentum $l_0$  when the electron exits the tunnelling barrier. In the tunnelling limit  $l_{0}\rightarrow 0$, since when the electron exits the tunnelling barrier its velocity is parallel to electron displacement. Thus, for typical field strength $F=0.06$ a.u. and $I_{p}=0.5$ a.u. the phase difference $\xi_{SO}\sim0.21F^2/(c^2I_p^{5/2})\sim 2.3\times10^{-7}$~rad. and is completely negligible.

\section{Rotation of the hole spin in strong field ionisation: Gedanken experiment in Kr atom} \label{app:iii}

Consider Kr atom in its ground state. There is no spin-orbit interaction in the ground state of the neutral Kr: the $P$-shell is filled by 6 $p$-electrons, with equal number of $p^{-}$ and $p^{+}$ electrons `rotating' in opposite directions. Ionisation by strong, circularly polarised IR laser field breaks the balance between $p^{-}$ and $p^{+}$ electrons \cite{Barth2011} and 
starts the spin-orbit Larmor clock. Intense right-circularly polarised IR pulse prefers to remove the $p^{-}$ electron \cite{Barth2011,Herath2012}, i.e. $m_l=-1$. Let us set the initial spin state to be $\vert s_{\rm in} \rangle=\alpha\vert -1/2 \rangle + e^{i\phi}\beta\vert 1/2 \rangle$, where $\alpha, \beta$ are real numbers and the phase $\phi$ characterises the initial orientation of the spin. Once the $p^{-}$ electron is removed, the quantum state of the core acquires uncompensated angular momentum, with the hole created with  $M_{L}=-1$ and uncompensated spin. The spin state of the hole is $\vert s_{\rm in} \rangle=\alpha\vert -1/2 \rangle + e^{i\phi}\beta\vert 1/2 \rangle$, since the spins and the angular momenta of the hole and the electron are the same at the moment of separation. As this state is not an eigenstate of the Hamiltonian, the hole spin starts to precess.

We shall now calculate the angle of rotation of the hole spin. The final spin
state for the fixed orientation of the final orbital momentum $M_L=-1$
is $\vert s_{\text{fin}} \rangle=a_{\downarrow}\alpha \vert -1/2 \rangle + a_{\uparrow} e^{i\phi}\beta \vert 1/2 \rangle$, where the $a_{\uparrow}$ and $a_{\downarrow}$ are the strong field ionisation amplitudes for the spin-up and spin-down ionisation pathways. We first specify our notations and introduce the ionisation amplitude $T^{-}(I_p)$ corresponding to the removal of $p^{-}$ electron in the absence of the spin-orbit splitting of the core state. Here $I_p$ is the ionisation potential. The amplitudes $T_3$ and $T_1$, which include the spin-orbit splitting, are proportional to $T^{-}(I_p)$: $T_3^-\propto T^{-}(I_p)$ and $T_1^-\propto T^{-}(I_p+\Delta E_{SO})$, and they
correspond to the removal of the $p^{-}$ electron \cite{Barth2011,spin_currents}. 

Full ionisation amplitudes into the hole states $p_{J,M_J}$ include the projections $\langle LM_L,SM_S \vert JM_J \rangle$ given by the Clebsch-Gordan coefficients, $C_{LM_{L},\frac{1}{2}M_{S}}^{JM_{J}}$, with $M_L=1$.  Taking these projections into account, we find that the amplitude of ionisation into the hole state $J=3/2$, $M_J=3/2$ is $T_3^{-}$. The amplitude of ionisation into the state $J=3/2$, $M_J=1/2$ is $\frac{1}{\sqrt{3}}T_3^{-}$, the amplitude of ionisation into the state $J=1/2$, $M_J=1/2$ is $-\sqrt{\frac{2}{3}}T_3^{-}$. 
Now, we project these states back onto the $L,M_L,S,M_S$ basis to find $a_{\uparrow}$ and
$a_{\downarrow}$. 
This yields the amplitude to find the hole angular momentum $M_L=-1$ and $M_S=-1/2$ at time $t$  $a_{\downarrow}=T_3^{-}e^{-iE_{3/2}t}$, while the amplitude to find 
the hole angular momentum $M_L=-1$ and $M_S=1/2$ at the time $t$ is $a_{\uparrow}=\frac{1}{3}\left(2T_1^{-}e^{-iE_{1/2}t}+T_3^{-}e^{-iE_{3/2}t} \right)$. 
Here $E_{3/2}$ is the energy of the ground state, $\vert E_{1/2}\vert=\vert E_{3/2}\vert +\Delta E_{SO}$.

To establish the rotation angle (the phase of the double arm of the interferometer relative to the phase of the single arm (see Fig. 2(b)); the single arm now corresponds to spin-down pathway, while the double arm now corresponds to spin-up pathway) we need to find $\arg[a_{\downarrow}a^*_{\uparrow}]$:
\begin{multline}
a_{\downarrow}a^*_{\uparrow}=\frac{1}{3}\left(|T_3^{-}|^2+2|T_1^{-}||T_3^{-}|(\cos(\phi^{T}_3-\phi^{T}_1 + \right.\\\left.\Delta E_{SO}t)+i\sin(\phi^{T}_3-\phi^{T}_1+\Delta E_{SO}t)\right),
\end{multline}
\begin{equation}
\arg\left[T_3^{-}\left(\frac{1}{3}(T_3^{-*}+2T_1^{-*})\right)\right]=\arctan\frac{2|T_1^{-}||T_3^{-}|\sin(\phi^{T}_3-\phi^{T}_1+\Delta E_{SO}t)}{|T_3^{-}|^2+2|T_1^{-}||T_3^{-}|(\cos(\phi^{T}_3-\phi^{T}_1+\Delta E_{SO}t)}. \label{argT}
\end{equation}
Equation~\eref{argT} yields:
\begin{equation}
\tan\Delta\phi_{SO}=\frac{\sin(\Delta E_{SO}t-\Delta\phi_{13})}{0.5|T_3^{-}|/|T_1^{-}|+\cos(\Delta E_{SO}t-\Delta \phi_{13})}, \label{Eq:Larmor1S}
\end{equation}
where the phase difference $\Delta\phi_{13}$ is defined as $\Delta\phi_{13}=\phi^{T}_1-\phi^{T}_3$.

\section{Core potentials in two different ionisation channels} \label{app:iv}

To illustrate the effect of electron-hole correlations on definition and measurement of time, we consider the contribution of 
the channel specific core potential $V_{LJM_J}(\mathbf{r})$, that arises from the Coulomb interaction between the electron and the core. This potential has the following form:
\begin{multline}	
V_{LJM_J}(\mathbf{r}) = \int_{}^{}d\mathbf{r}'\frac{\rho_{\tr{ion}}(\mathbf{r}')}{\|\mathbf{r}-\mathbf{r}'\|} = \int_{}^{}d\mathbf{r}'\frac{1}{\|\mathbf{r}-\mathbf{r}'\|}\langle\epsilon JM_J|\mathbf{r}'\rangle\langle\mathbf{r}'|\epsilon JM_J\rangle =\\
\sum_{\substack{M_{L},M_{L}',M_S,M_S'}}C_{LM_{L}',\frac{1}{2}M_S'}^{JM_J}C_{LM_{L},\frac{1}{2}M_S}^{JM_J}\left\langle\frac{1}{2}M_S'\right.\left|\frac{1}{2}M_S\right\rangle\int_{}^{}d\Omega\,Y_{LM_{L}'}^*(\theta',\phi')Y_{LM_{L}}(\theta',\phi') \times\\
\sum_{L_1=0}^{\infty}P_{L_1}(\cos\beta)\left[\int_{0}^{r}dr'r'^2\frac{r'^{L_1}}{r^{L_1+1}}\left|R(\epsilon JM_J;r')\right|^2 + \int_{r}^{\infty}dr'r'^2\frac{r^{L_1}}{r'^{L_1+1}}\left|R(\epsilon JM_J;r')\right|^2\right],
\end{multline}
where, $L = J \pm 1/2$, is the orbital angular momentum fixed for a given spin-orbital, $\beta$ is the solid angle between the vectors $\mathbf{r}$ and $\mathbf{r}'$, and can be written as $\cos\beta = \hat{\mathbf{r}}\cdot\hat{\mathbf{r}}'$, $\epsilon$ represents the effective principle quantum number corresponding to the energy of the spin-orbital under consideration, and $R(\epsilon JM_{J};r)$ is the radial part of the wavefunction associated to the said spin-orbital.

Including all terms together, we have:
\begin{multline}
V_{LJM_J}(\mathbf{r}) = \sum_{\substack{M_{L},M_{L}'\\M_S}}C_{LM_{L},\frac{1}{2}M_S}^{JM_J}C_{LM_{L}',\frac{1}{2}M_S}^{JM_J}\sum_{L_1=0}^{\infty}\frac{4\pi}{2L_1+1}\sum_{M_{L_1}=-L_1}^{L_1}Y_{L_1 M_{L_1}}^*(\theta,\phi)\langle R_{L_1}\rangle \times \\
\int_{}^{}d\Omega\,Y_{LM_{L}'}^*(\theta',\phi')Y_{LM_{L}}(\theta',\phi')Y_{L_1 M_{L_1}}(\theta',\phi').
\end{multline}
Here $\phi$ is the angle in polarisation plane, $\theta$ is the angle calculated from the laser propagation direction, and $\langle R_{L_{1}} \rangle$ is the expectation value of the radial component, as calculated using the Roothaan-Hartree-Fock (RHF) orbitals, defined as:
\begin{equation}
R_{L}(r) = \sum_{p,q}c_{i_{p}}c_{i_{q}}\left[\frac{1}{r^{L+1}}\gamma((\kappa_{i_{p}} + \kappa_{i_{q}})r) + r^{L}\Gamma((\kappa_{i_{p}} + \kappa_{i_{q}})r)\right],
\end{equation}
where, $c_{i_{p}}$, $c_{i_{q}}$ are the coefficients for the Slater-Type Orbitals (STO) and $i_{p}$, $i_{q}$ the corresponding indices defining the nodes in the wavefunction under consideration, used for the RHF calculations \cite{rhf1993}, and $\gamma$ is the lower, whereas $\Gamma$ is the upper incomplete-gamma function.
Taking into account Wigner $3j$-coefficients from the integral:
\begin{multline}
\int_{}^{}d\Omega\,Y_{LM_{L}'}^*(\theta',\phi')Y_{LM_{L}}(\theta',\phi')Y_{L_1 M_{L_1}}(\cos\theta') = (-1)^{M_{L}'}(2L+1)\sqrt{\frac{2L_1+1}{4\pi}} \times \\
\begin{pmatrix}
L & L_1 & L \\
M_{L} & M_{L_1} & -M_{L}' 
\end{pmatrix}
\begin{pmatrix}
L & L_1 & L\\
0 & 0 & 0
\end{pmatrix},
\end{multline}
we obtain the selection rules. For $L_1$, the selection rules are: (a) $2L+L_1$ is even (so only $L_1$ even are allowed in the summation over $L_1$) and (b) the triangle inequality $|L - L_1| \leq L \leq L + L_1$
which gives $0 \leq L_1\leq 2L$. For all other cases the integral is zero, and $M_{L}' = M_{L} + M_{L_1}$.

Taking $L_1 = 2L'$, the expression for $V_{LJM_J}$ is:
\begin{multline}
V_{LJM_J}(\mathbf{r}) = (2L+1)\sum_{M_{L},M_S}\sum_{L'=0}^{L}\sum_{M_{2L'}=-2L'}^{2L'}(-1)^{M_{L}+M_{2L'}}C_{L{M_{L}+M_{2L'}},\frac{1}{2}M_S}^{JM_J}C_{L{M_{L}},\frac{1}{2}M_S}^{JM_J} \times \\
\sqrt{\frac{4\pi}{4L'+1}}
\begin{pmatrix}
L & 2L' & L\\
M_{L} & M_{2L'} & -M_{L}'
\end{pmatrix}
\begin{pmatrix}
L & 2L' & L\\
0 & 0 & 0
\end{pmatrix}
Y_{2L'M_{2L'}}^*(\theta,\phi)\langle R_{2L'}\rangle.
\end{multline}
From the Clebsch-Gordan coefficients, we have two conditions on $M_{L}$ and $M_S$ for a given $M_{J}$:
\begin{gather}
M_{L} + M_{2L'} + M_S = M_{J},\\
M_{L} + M_S = M_{J},
\end{gather}
which can only be possible if $M_{2L'} = 0$. The final expression is:
\begin{multline}
V_{LJM_J}(\mathbf{r}) = (2L+1)\sum_{M_{L},M_S}\sum_{L'=0}^{L}(-1)^{M_{L}}\left\vert C_{LM_{L},\frac{1}{2}M_S}^{JM_J} \right\vert^2\sqrt{\frac{4\pi}{4L'+1}}
\begin{pmatrix}
L & 2L' & L\\
M_{L} & 0 & -M_{L}
\end{pmatrix} \times \\
\begin{pmatrix}
L & 2L' & L\\
0 & 0 & 0
\end{pmatrix}
Y_{2L'0}^*(\theta,\phi)\langle R_{2L'}\rangle
\end{multline}
Using the definition of $Y_{2L'0}$, we can simplify further to give:
\begin{multline}
V_{LJM_J}(\mathbf{r}) = (2L+1)\sum_{M_{L},M_S}\sum_{L'=0}^{L}(-1)^{M_{L}}\left|C_{LM_{L},\frac{1}{2}M_S}^{JM_J}\right|^2
\begin{pmatrix}
L & 2L' & L\\
M_{L} & 0 & -M_{L}
\end{pmatrix} \times \\
\begin{pmatrix}
L & 2L' & L\\
0 & 0 & 0
\end{pmatrix}P_{2L'}(\cos\theta)\langle R_{2L'}\rangle.
\end{multline}
Note that $L' = 0$ corresponds to Coulomb potential, common in both channels. Consider the case when ionisation liberates the $p^{+}$ electron ($L=1$) populating the hole states $J = 3/2,1/2$ and $M_J = 1/2$ (the result for $p^-$ is the same). For the calculation of the difference between two core potentials we use the same trajectory with averaged $I_p$. The corrections associated with the difference in the trajectories are of higher order and are not included here.

The difference in core potentials for this trajectory is:
\begin{multline}
V_{1,3/2,1/2}(\mathbf{r}) - V_{1,1/2,1/2}(\mathbf{r}) = 3\sum_{L'=0}^{1}
\begin{pmatrix}
1 & 2L' & 1\\
0 & 0 & 0
\end{pmatrix}
P_{2L'}(\cos\theta)\langle R_{2L'}\rangle \times \\
\left[\sum_{M_{L},M_S}(-1)^{M_{L}}\left\vert C_{1M_{L},\frac{1}{2}M_S}^{3/2\,1/2} \right\vert^2
\begin{pmatrix}
1 & 2L' & 1\\
M_{L} & 0 & -M_{L}
\end{pmatrix} \right. - \\
\left. \sum_{M_{L},M_S}(-1)^{M_{L}}\left|C_{1M_{L},\frac{1}{2}M_S}^{1/2\,1/2}\right|^2
\begin{pmatrix}
1 & 2L' & 1\\
M_{L} & 0 & -M_{L}
\end{pmatrix}\right].
\end{multline}
As expected for the common Coulomb potential, the difference for $L' = 0$ is zero:
\begin{multline}
\sum_{M_{L},M_S}(-1)^{M_{L}}
\begin{pmatrix}
1 & 0 & 1\\
M_{L} & 0 & -M_{L}
\end{pmatrix}
\left[\left|C_{1M_{L},\frac{1}{2}M_S}^{3/2\,1/2}\right|^2 - \left\vert C_{1M_{L},\frac{1}{2}M_S}^{1/2\,1/2} \right\vert^2\right] =\\
\begin{pmatrix}
1 & 0 & 1\\
0 & 0 & 0
\end{pmatrix}\left[\frac{1}{3}\right] +(-1)
\begin{pmatrix}
1 & 0 & 1\\
1 & 0 & -1
\end{pmatrix}\left[-\frac{1}{3}\right] = -\sqrt{\frac{1}{3}}\frac{1}{3} + \sqrt{\frac{1}{3}}\frac{1}{3} = 0.
\end{multline}
The only term left is the one corresponding to $L' = 1$, which gives
\begin{multline}
V_{1,3/2,1/2}(\mathbf{r}) - V_{1,1/2,1/2}(\mathbf{r}) = 3
\begin{pmatrix}
1 & 2 & 1\\
0 & 0 & 0
\end{pmatrix}P_{2}(\cos\theta)\langle R_{2}\rangle \times \\
\sum_{M_{L},M_S}(-1)^{M_{L}}
\begin{pmatrix}
1 & 2 & 1\\
M_{L} & 0 & -M_{L}
\end{pmatrix}
\left[\left\vert C_{1M_{L},\frac{1}{2}M_S}^{3/2\,1/2} \right\vert^2 -
\left\vert C_{1M_{L},\frac{1}{2}M_S}^{1/2\,1/2} \right\vert^2
\right] =\\
3\sqrt{\frac{2}{15}}P_2(\cos\theta)\langle R_{2}\rangle\left[\sqrt{\frac{2}{15}}\frac{1}{3} + (-1)\sqrt{\frac{1}{30}}\left(-\frac{1}{3}\right)\right] = \frac{1}{5}P_2(\cos\theta)\langle R_2\rangle = -\frac{\langle R_2\rangle}{10}, \label{EqApp:hart_pot1}
\end{multline}
since for $\theta = \pi/2$, $P_2 = -1/2$. The expression for $\langle R_2\rangle$ is:
\begin{equation}
R_2 = \int_{0}^{r}dr'r'^2\frac{r'^2}{r^3}|R(\epsilon L;r')|^2 + \int_{r}^{\infty}dr'r'^2\frac{r^2}{r'^3}|R(\epsilon L;r')|^2 = \frac{1}{r^3}\langle r'^4\rangle_0^r + r^2\left\langle\frac{1}{r'}\right\rangle_r^\infty, \label{EqApp:hart_pot2}
\end{equation}
which can be found from the incomplete gamma functions. The difference between the two core potentials is: $V_{1\,3/2\,1/2}(\mathbf{r}) - V_{1\,1/2\,1/2}(\mathbf{r})\simeq-4.444/(10r^3)$, since $\langle R_2\rangle=4.444$ a.u. for Kr \cite{Radzig} and the contribution of the second term in Eq.~\eref{EqApp:hart_pot2} vanishes for $r\rightarrow\infty$. To calculate the respective relative phase $\Delta\phi^d_{13}$, we use Eq.~\eref{eq:WC} and substitute the difference in short range core potentials given above.

\section{Pump-probe signal: the details of derivation} \label{app:v}

The goal of this section is to derive population in the final
$S$-state of the Kr ion at the end of the pump-probe experiment, see Eq.~\eref{Eq:Larmor1a} of the main text.

For a laboratory experiment, we need two requirements. First, we want to turn on and turn off the clock on demand, i.e. we need to stop the rotation of the core spin on demand. Second, we would like to measure the phase $\Delta \phi_{13}$ directly. The second condition is satisfied automatically, since the initial superposition of spin up and spin down states is incoherent and therefore the single arm of the interferometer (in Fig.~\ref{figure2}(b)) will not interfere with the double arm in a real experiment. Thus, the laboratory experiment will only record the interference in the double arm, and the single arm will give background. To start the clock, we apply a nearly single-cycle right circularly polarised IR pulse to create a $p$-hole. To stop the clock, we apply a left circularly polarised laser field to induce a transition from the $s$-shell of the Kr atom, filling the $M_L=-1$ hole in the $p$-shell and leaving the hole in $s$-orbital. There is no angular momentum in the $s$-hole, and there is no SO interaction. Thus, the left-circular probe stops the clock that was started by the right circular pump.

For a fixed final state of the continuum electron, characterised by momentum $\mathbf{p}$ at the detector, the population $S=\left\vert \sigma_{1/2,M_L=0} \right\vert^2 + \left\vert \sigma_{-1/2,M_L=0} \right\vert^2$ in the final $s$-state can be obtained using the following equation:
\begin{equation}
\sigma_{M_S,M_L=0} = \int_{}^{} dt\,\left\langle \Psi_{\text{fin}}(t) \middle\vert \widehat{d} \middle\vert \Psi_{\text{ion}}(t) \right\rangle E_{\text{asec}}(t).
\end{equation}
where $\Psi_{\text{ion}} = \sum_{J,M_J}a_{JM_{J}}\psi_{JM_{J}}e^{-iE_{J}t}$ is the coherent superposition of the two ionic states, created after ionisation, for a given final momentum of the electron at the detector. Here $a_{JM_J}$ is the complex amplitude of ionisation into hole state $\vert J, M_{J} \rangle$. The wavefunction $\Psi_{\text{fin}}(t)=\psi_{\text{S}}e^{-iE_St}$ represents the final $S$-state of the ion.

Following Fig. 4(b), we write  one particle spin-orbitals participating in the one-electron transition from the inner valence $s$-orbital to the outer-valence $J,M_J$ orbital  $\psi_{JM_J}$ and $\psi_{\text{S}}$ as a product of angular and radial wave-functions: $\psi_{JM_J} =  \psi_{J}(r)  \vert JM_J \rangle$, $\psi_{S} = \psi_{S}(r) \vert LM_{L},SM_{S} \rangle$. Taking into account that $L=0$, $M_L=0$, $S=1/2$ in the final state, we obtain: $\psi_{S} = \psi_{S}(r)\vert 0\,0,1/2\,M_S \rangle$.

The dipole operator can be factorised into the radial and angular parts, $\widehat{d} = \widehat{r}\,\widehat{\Xi}_{\xi}$, where $\xi=1$ corresponds to the right polarised pulse, $\xi = -1$ corresponds to the left polarised pulse: $\widehat{\Xi}_{\xi} = d_x + i\xi d_y$. Evaluating the integral over $t$, we rewrite the equation in equivalent form:

\begin{multline}
\sigma_{M_S,M_L=0}=\sum_{J,M_J,M_L,M_S'}{F_{\omega}}(E_S-E_J) a_{JM_J}d_J \\
\times \left\langle JM_J  \middle\vert 1\,M_L,1/2\,M_S' \right\rangle \\ \times \left\langle 1\,M_L,1/2\,M_S'\middle\vert \widehat{\Xi}_{\xi} \middle\vert  0\,0,1/2\,M_S \right\rangle.
\end{multline}
Here, ${F_{\omega}}(E_S-E_J)$ is the Fourier image of the probe pulse $F_{\text{asec}}(t)$ taken at the transition frequency. The real-valued radial matrix element $d_J = \left\langle \psi_J(r) \middle\vert \widehat{r} \middle\vert \psi_S(r) \right\rangle$ describes the electron transition from the inner-valence state $S$ to the final sate $\vert JM_J\rangle$, leaving the hole in $S$-orbital. For the left circularly polarised field, the angular part of the dipole operator $\left\langle 1\,M_L=-1,1/2\,M_S'\middle\vert \widehat{\Xi}_{-1} \middle\vert  0\,0,1/2\,M_S \right\rangle                                                                                                                                                                                                                                                                                                                                                                                                                                                                                                                                                                                                                                                                                                                                                                                                                                                                                                                                                                                                                                                                                                                                                                                                                                                                                                                                                                                                                                                                                                                                                                                                                                                                                                                                                                                                                                                                                                                                                                                                                                                                                                                                                                                                        =\delta_{M_S,M_S'}\sqrt{N}$, $N=2/3$. Thus, we obtain

\begin{equation}
\sigma_{M_S,M_L=0}=\sum_{J,M_J}\sqrt{N}d_{J}{F_{\omega}}(E_S-E_J) a_{JM_J} \left\langle JM_J  \middle\vert 1\,-1,1/2\,M_S \right\rangle .\end{equation}
The Clebsch-Gordan coefficients $C_{L=1 M_L=-1,S=1/2M_S}^{JM_J}=\left\langle JM_J  \middle\vert 1\,-1,1/2\,M_S \right\rangle$ are equal to: $C_{1,-1,1/2,-1/2}^{3/2,-3/2}=1$, $C_{1,-1,1/2,1/2}^{3/2,-1/2}=1/\sqrt{3}$, $C_{1,-1,1/2,1/2}^{1/2,-1/2}=-\sqrt{2/3}$. 


Left polarised pulse promotes s-electron to $M_L=-1$ hole, right polarised pulse to $M_L=1$ hole, linearly polarised pulse to $M_L=0$ hole. Thus, if right circularly polarised pulse is used as a pump, left circularly polarised probe will probe ionisation of $p^{-}$ electron, whereas right circularly polarised probe will probe ionisation of $p^{+}$ electron. Linearly polarised probe will probe both $p^{+}$ and $p^{-}$ pathways at the same time.
For the left-circularly polarised probe we obtain:
\begin{equation}
\left\vert \sigma_{1/2,0} \right\vert^2 = \frac{N}{9}d_{3/2}^2\left\vert T_3^{-} \right\vert^2 \left\vert F_{\omega}(\Omega_{3}) \right\vert^2,
\end{equation}
and
\begin{multline}
\left\vert \sigma_{-1/2,0} \right\vert^2 = \frac{N}{9}d_{3/2}^2\left\vert T_3^{-} \right\vert^2\left\vert F_{\omega}(\Omega_3) \right\vert^2+\frac{4N}{9}d_{1/2}^2\left\vert T_1^{-} \right\vert^2\left\vert F_{\omega}(\Omega_1) \right\vert^2 + \\ \frac{4N}{9}d_{1/2}d_{3/2} \left\vert T_{1}^{-} \right\vert\left\vert T_3^{-} \right\vert \left\vert F_{\omega}(\Omega_3)\right\vert \left\vert F_{\omega}(\Omega_1)\right\vert\cos(\Delta E_{SO}\tau-\Delta \phi_{13}),
\end{multline}
where $\tau$ is the time of arrival of the attosecond pulse, $\Omega_{3} = E_S-E_{3/2}$, $\Omega_{1}=E_S-E_{1/2}$, $E_S$ is the energy of the final $S$-state. Transform limited attosecond pulse is assumed for this calculation. The population in $S$-state is $S = \left\vert \sigma_{1/2,M_L=0} \right\vert^2 + \left\vert \sigma_{-1/2,M_L=0} \right\vert^2$, and can be measured by transient absorption of the XUV probe.



\section*{References}
\providecommand{\newblock}{}


\end{document}